\documentstyle[amsfonts,amscd,12pt]{amsart} 
\theoremstyle{definition}
\theoremstyle{remark} \numberwithin{equation}{section}

\begin{document} 
\title[K\"ahler polarisation
and Wick quantisation]{K\"ahler polarisation
and Wick quantisation \\ of Hamiltonian systems \\
subject to second class constraints}
\author{S.L.Lyakhovich ${}^{\small a, (*)}$}
\author{A.A.Sharapov ${}^{\small b}$}
\thanks{ (*) On leave of absence from Tomsk
State University, Russia}
\dedicatory{$a$ Institute for Theoretical Physics, Chalmers University
of Technology, G\"oteborg, Sweden
 \\ $b$ Department of Theoretical Physics, Physics Faculty, Tomsk State
University, Russia} \maketitle

\begin{abstract} The necessary and sufficient conditions are established for the
second-class constraint surface to be (an almost) K\"ahler manifold.  The
deformation quantisation for such systems is scetched resulting in the Wick-type
symbols for the respective Dirac brackets.  \end{abstract}








\section{Introduction}

The progress of the quantum field theory was always related, directly or
indirectly, with in-depth study of quantisation methods.  Recent years have
brought the explosive developments in the Deformation Quantization theory (see
\cite{St} for review and further references).  By the optimistic expectations an
appropriate adaptation of this method to the field theory would help to resolve
some old problems concerning the non-perturbative quantisation of essentially
non-linear models (like the non-linear sigma-models on a homogeneous background
and, in particular, strings on AdS space), when any linear approximation can
violate fundamental symmetries of the theory.  To provide a natural frame for a
particle interpretation of the fields quantised, the Wick-type deformation
quantisation seems to be most appropriate (see \cite{BW}, \cite{DLS}, \cite{KS}
an references therein).

Any actual application of these methods to the field theory should also be
supplemented by a proper account of the Hamiltonian constraints, which are
characteristic for all practically interesting field theory models.  The first
step in this direction was done in \cite{BGL}, where the BRST version \cite{GL}
of the Fedosov deformation quantization \cite{Fedosov} was extended to the
second-class constraint systems.  The next step would consist in generalising
the construction of the paper \cite{BGL} (giving the Weyl type symbols for the
second class systems) to the Wick symbol calculus on a non-linear phase-space
subject to the Hamiltonian constraints.  As the Wick structure is usually
inherited from the K\"ahler structure, one naturally comes to consideration of the
Hamiltonian reduction on the (almost-)K\"ahler manifolds.  In this note we
study, at first, the question of the classical Hamiltonian reduction by second
class-constraints and examine the compatibility of this reduction with the
K\"ahler structure of the enveloping manifold.  The extension of these results
to the quantum case is briefly discussed in the concluding section, more details
will be given elsewhere.

\section{Almost-K\"ahler manifolds}

In this section we briefly recall some basic definitions and facts concerning
the geometry of almost-K\"ahler manifolds.  For more details, see, for example,
\cite{Yano}.

The almost-K\"ahler manifold $(M,J,\omega )$, is a real $2n$-dimensional
manifold $M$ together with an almost-complex structure $J$ and a symplectic form
$\omega $ which are compatible in the following sense:
\begin{equation} \omega (JX,JY)=\omega (X,Y)
\end{equation} for any vector fields $X,Y$.  In other
words, the smooth field of authomorphisms
\begin{equation} J:TM\rightarrow
TM,\qquad J^2=-1 \label{can}
\end{equation}
is a canonical transformation of the
tangent bundle w.r.t.  symplectic structure $\omega $.
Then $g(X,Y)=\omega (JX,Y)$ is $J$-invariant (pseudo-)Riemannian metric on $M$,
\begin{equation}
g(JX,JY)=g(X,Y).  \end{equation}
The almost-complex structure $J$ splits the
complexified tangent bundle $T^{\Bbb C}M$ onto two transverse mutually
conjugated sub-bundles:  $T^{\Bbb C} M=T^{(1,0)}M \oplus T^{(0,1)}M$, such
that
\begin{equation} J_pX=iX,\qquad \forall X\in T_p^{(1,0)}M, \label{x}
\end{equation}
\begin{equation} J_pY=-iY,\qquad \forall Y\in T_p^{(0,1)}M,
\label{y} \end{equation}
for every $p\in M$.  In the natural frames
$\partial_i=\partial /\partial x^i,$ $dx^j$ associated to local coordinates
$(x^i)$ on
$M$ we have $ J_j^i=J(dx^i,\partial _j)$,
$\omega_{ij}=\omega (\partial_i,\partial_j)$,
$g_{ij}=g(\partial_i,\partial_j),$ and
\begin{equation}
J_j^i=g^{ik}\omega_{kj}=g_{jk}\omega^{ki},
\end{equation}
where $(g^{ik})$
and
$(\omega^{ik})$ are the inverse matrices to $(g_{ij})$ and $(\omega_{ij}),$
respectively.  It is well known that any symplectic manifold $(M,\omega )$
admits a compatible almost-complex structure $J$, which turns it to an
almost-K\"ahler manifold.

Alternatively, the almost-K\"ahler manifold can be defined as a pair
$(M,\Lambda )$ in which $M$ is $2n$-dimensional real manifold equipped with a
degenerate Hermitian form $\Lambda ,$ such that
\begin{equation} {\rm rank}\Lambda =\frac 12\dim M=n,
\label{rank} \end{equation}
\begin{equation}
\det ({\rm Im}\Lambda )\neq 0.  \end{equation}
The equivalence of both the
definitions is set by the formula
\begin{equation}
\Lambda_{jk}=g_{jk}+i\omega_{jk}.
\label{lambda}
\end{equation}
The vector fields of type $(1,0)$ or $(0,1)$ w.r.t.  the
almost-complex structure $J$, being defined by relations (\ref{x}), (\ref{y}),
are nothing but the right/left null-vectors for the form $\Lambda $.

Now let $\nabla $ be the unique torsion-free connection respecting metric $g$.
This connection is known to respect the symplectic structure $\omega $ whenever
$(M,J,\omega )$ is the K\"ahler manifold (equivalently, iff $J$ is integrable),
so that tensor
\begin{equation} T_{jk}^i=\omega^{in}\nabla_n\omega_{jk},
\label{LT} \end{equation}
vanishes.  In general case one can always define a new
affine connection $ \overline{\nabla }$, which already respects both the metric
and symplectic structures, by adding an appropriate torsion.  The canonical
possibility is to put $\overline{\nabla } = \nabla +T$.  Indeed,
\begin{equation}
\overline{\nabla }_i \omega_{jk}=\nabla _i \omega_{jk}-T_{ij}^n\omega_{nk}-
T_{ik}^n \omega_{nj}=
\end{equation}
\begin{equation*} =\nabla _i
\omega_{jk}+\nabla_j\omega_{ki}+\nabla_k \omega_{ij}=(d\omega )_{ijk}=0.
\end{equation*} This choice for $\overline{\nabla }$ is known as the
Lichnerowicz connection \cite{Lichne} of an almost-K\"ahler manifold.
Thus, the
existence of a torsion-free connection respecting Hermitian form $\Lambda $ is
the necessary and sufficient condition for the almost-K\"ahler manifold $M$ to
be a K\"ahler one.  Introduce the Hermitian form
$\Lambda^{ik}=g^{ik}+i\omega^{ik}$ on the complexified cotangent bundle of $M$.  In view of the identity
$\Lambda^{ij}\Lambda_{jk}=0$ and the rank condition ({\ref{rank}}) any vector
field $X$ of type $(0,1)$ can be (non-uniquely) represented as
\begin{equation}
X=Z_j\Lambda^{ji}\partial_i \label{v1}
\end{equation} for some one-form
$Z=Z_i(x)dx^i$.  Commuting two vector fields of the form ({\ref{v1}}) and
requiring the result to annihilate the form $\Lambda $
(\ref{lambda}) on the left we get the
integrability condition for the distribution $T^{(0,1)}M$:
\begin{equation}
(\Lambda^{in}\partial_n\Lambda^{jk}-
\Lambda^{in}\partial_n\Lambda^{jk})\Lambda_{km}=0
\end{equation}
Similarly, the subbundle $T^{(1,0)}M$ is
integrable iff
\begin{equation}
(\Lambda^{ni}\partial_n\Lambda ^{jk}-\Lambda^{ni}\partial_n\Lambda^{jk})
\Lambda_{mk}=0.
\end{equation}
(Here we use of the identity $\Lambda_{ki}\Lambda^{ij}=0.$)

Surprisingly, the Lichnerowicz torsion $T$ being associated to the metric $g$
and the symplectic form $\omega $ can be written in terms of the almost-complex
structure $J$ along.  After some algebra one may find that \begin{equation}
T_{jk}^i=-\frac 14N_{jk}^i,\quad N_{jk}^i=J_k^n\partial_n J_j^i-J_j^n
\partial_nJ_k^i+J_n^i\partial _jJ_k^n-J_n^i\partial _kJ_j^n
\end{equation} The tensor
$N$ is known as Nijenhuis tensor of an almost-complex structure $J $.  It is
$N\neq 0$ which is the only obstruction for the structure $J$ to be integrable.

>From the view point of symplectic geometry, the integrable
holomorphic/anti-holomorphic distributions $T^{(1,0)}M$ and $T^{(0,1)}M$ define
a pair of transverse Lagrangian polarizations of $M$, i.e.
$\omega |_{T^{(1,0)}M}=\omega |_{T^{(0,1)}M}=0$.
The existence of such polarizations is
of primary importance for the physical applications as it makes possible to
define the notion of {\it physical state }for a quantum-mechanical system, at
least in the framework of geometric quantization \cite{W}.

\section{Hamiltonian reduction by second-class constraints.}

Consider $(M,J,\omega )$ as the phase-space of a mechanical system with the
Poisson bracket
\begin{equation*} \{f,g\}=\omega^{ij}\partial_i f\partial_j g, \qquad
f,g\in C^\infty (M), \end{equation*} and let
$\theta_a$ be a set of
$2m$, ($m<n$) smooth independent functions on $M$ such that the matrix of
Poisson brackets
\begin{equation}
\omega_{\alpha \beta }=\{\theta_\alpha ,\theta_\beta \}
\label{mat}
\end{equation}
is non-degenerate on the whole phase space,
$det \, \omega_{\alpha\beta} \neq 0$.
Then equations
\begin{equation}
\theta_\alpha =0,\qquad \alpha =1,...,2m,
\label{constr}
\end{equation}
called second-class constraints, extract $2(n-m)$-dimensional
smooth surface $\Sigma \subset M$.
It is the surface where the dynamics of the
mechanical system is assumed
to evolve\footnote{ Note, that the usual
definition of the second-class constraints requires the invertibility of the
matrix (\ref{mat}) only on the constraint surface (\ref {constr}) and hence, in
some it's tubular neighborhood.  However, it is not clear at the moment whether
it is possible to perform the consistent (algebraic, geometric, deformation,
...)  quantisation under such a weakened condition without explicit solving the
constraints.}.
As the matrix (\ref{mat}) is nowhere degenerate, the pair
$(\Sigma ,\omega |_\Sigma )$ is a symplectic manifold again,
$\omega |_\Sigma $
is a restriction of two-form $\omega $ onto surface $\Sigma $.
Denote by
$\{\cdot ,\cdot \}_\Sigma $
the corresponding Poisson bracket.  According to the
general theory of second-class constraint systems the Poisson bracket
$\{\cdot,\cdot \}_\Sigma $ can be extended to a bracket on the whole manifold
$M$.  This extension, known as Dirac bracket, is given by
\begin{eqnarray}
\{f,g\}_D
&=&\{f,g\}-\{f,\theta_\alpha \}\omega^{\alpha \beta }\{\theta_\beta
,g\}=\widetilde{\omega }^{ij}\partial_if\partial_jg, \notag \\ &&
\label{DB}
\\ \widetilde{\omega }^{ij} &=&\omega ^{ij}-
\omega^{in}\partial_n\theta_\alpha
\omega^{\alpha \beta }\partial_m\theta_\beta \omega ^{mj} \notag
\end{eqnarray}
where $(\omega^{\alpha \beta })$ is the inverse matrix for
$(\omega_{\alpha \beta })$.  Since the rank of Poisson bi-vector
$\widetilde{\omega }$ is constant, rel.  (\ref{DB}) defines a regular Poisson
bracket on $M$, having $ \theta $'s as the Casimir functions, i.e.
\begin{equation}
\{ f, \theta_\alpha \}_D=\{ \theta_\alpha ,f\}_D=0, \qquad
\forall f\in C^\infty (M).
\end{equation}
As any regular Poisson manifold,
$(M,\{\cdot ,\cdot \}_D)$ has a symplectic foliation:  the surfaces of constant
values of the Casimir functions are symplectic submanifolds with symplectic
structures induced by $\omega $.  In particular, the constraint surface
$\Sigma$ is a symplectic leaf corresponding to zero locus of the Casimir
functions.

We would like to emphasize that the Dirac bracket construction strongly depends
on the particular choice of the constraint functions, not just on the embedding
$ \Sigma \subset M$ itself.  Taking another constraint basis
\begin{equation}
\theta_\alpha \rightarrow \theta_\alpha ^{\prime }=
A_\alpha^\beta (x)\theta_\beta ,\qquad \det (A_\alpha^\beta )\neq 0,
\end{equation}
which defines the same submanifold $\Sigma $ by imposing the equations
$\theta^\prime_\alpha = 0$,
one gets a different extension for the Poisson bracket
$\{\cdot ,\cdot \}_\Sigma
$, which, however, coincides with (\ref{DB}) on the constraint surface
$\Sigma$.

An important advantage of the Dirac bracket as compared to the bracket $\{\cdot
,\cdot \}_\Sigma $ on the reduced phase space is that it allows to describe the
constrained dynamics without explicit solving constraints (the latter may be
rather problematic, especially in the field-theoretical models).  Another way to
deal with the second-class constraints (which probably is more suitable for the
subsequent quantisation) is to convert them into the first class ones by
extending the original phase space.  For this end consider the direct product
${\cal M}=M\times {\Bbb R}^{2m}$ equipped with the symplectic form \cite{BGL}
\begin{equation}
\Omega =\omega _{ij}dx^i\wedge dx^j+d \eta^\alpha \wedge
d\theta _\alpha , \end{equation} where $\eta^\alpha \, , \alpha =1, \ldots ,
2m$ are linear coordinates on ${\Bbb R}^{2m}$.  The corresponding Poisson
bracket $\{\cdot ,\cdot \}_{{\cal M}}$ is determined by the matrix inverse to
$\Omega $, in the local coordinates $(\xi ^A)=(x^i,\eta^\alpha )$ it reads
\begin{equation} \Omega ^{-1}=(\Omega ^{AB})=\left( \begin{array}{cc}
\widetilde{\omega }^{ij} & -\omega ^{il}\partial_l\theta _\gamma \omega
^{\gamma \beta } \\ \omega^{jl}\partial_l\theta_\gamma \omega^{\gamma \alpha
} & \omega^{\alpha \beta } \end{array} \right) \end{equation} As is seen the
upper left block of $\Omega ^{-1}$ is nothing but the Dirac bi-vector
(\ref{DB}).  So, the Poisson bracket $\{\cdot ,\cdot \}_{{\cal M}}$ coincides
with the Dirac one when is evaluated on $\eta $-independent functions
\footnote{This observation offers the most simple proof of the Jacobi identity
for the Dirac bracket:  it takes place because $\Omega $.  is closed.}.  In
particular, the constraints $\theta _\alpha $, considered as the functions on
the extended phase space ${\cal M}$, turn out to be in involution,
\begin{equation} \{\theta_\alpha ,\theta_\beta \}_{{\cal M}}=0 \end{equation}
The Hamiltonian reduction of $({\cal M},$ $\{\cdot ,\cdot \}_{{\cal M}})$ w.r.t.
the first-class constraints $\theta _\alpha $ immediately leads to the dynamics
on the constrained phase space $(\Sigma ,\{\cdot ,\cdot \}_\Sigma ).$ In so
doing, the equations \begin{equation} \eta^{\alpha}=0 \label{gauge}
\end{equation} may be thought of as admissible gauge fixing conditions for the
first-class constraints $\theta_\alpha ,$
\begin{equation} \{\theta_\alpha
,\eta^\beta \}=\delta_\alpha^\beta
\end{equation}
To put it differently, eqs.
(\ref{constr}), (\ref{gauge}), taken together,
define the second-class theory on
the extended phase space ${\cal M}$.
This extended constrained dynamics is
equivalent to the original second class constraint system on $M$.

Denote by $\Lambda |_\Sigma $ the restriction of the Hermitian form $\Lambda $
onto the constraint surface $\Sigma $.  The main question we address in this
section is in the following:  Given a second-class constraint system
$(M,\Lambda ,\theta _a),$
under which conditions the (almost)-K\"ahler structure
$\Lambda $ on $M$ induces a K\"ahler structure $\Lambda |_\Sigma$ on the
constraint surface $\Sigma :  \theta =0$ ?

To avoid having to deal with the degenerate structure of the Dirac bracket we
will perform all the calculations on the extended phase space ${\cal M}%
=M\times {\Bbb R}^{2m}$ and then reinterpret the results in terms of inner
geometry of $M$.  To begin with, we endow ${\cal M}$ with the
(pseudo-)Riemannian metric $G=G_{AB}d\xi{}^Ad\xi{}^B$ of the form \begin{equation}
(G_{AB})=\left( \begin{array}{cc} g_{ij} & \partial_i\theta_\alpha \\
\partial_j\theta & 0 \end{array} \right) \end{equation}
The inverse metric reads
\begin{equation} G^{-1}=(G^{AB})=
\left( \begin{array}{cc} \widetilde{{g}^{ij}} &
g^{il}\partial_l\theta_\gamma g^{\gamma \beta } \\ g^{jl}\partial_l
\theta_\gamma g^{\gamma \alpha } & -g^{\alpha \beta }
\end{array} \right) \label{BM}
\end{equation}
\begin{equation*}
\widetilde{{g}^{ij}}=g^{ij}-g^{in}\partial_n\theta_\alpha
g^{\alpha \beta }\partial_m\theta_\beta g^{mj}
\end{equation*}
As will be seen bellow, the direct product structure of the
extended phase-space, being considered in combination to the invariance of
$\Omega $ and $G$ under $\eta $-translations, allows to identify various
geometric structures on $M$ with certain blocks of geometric structures on
${\cal M}$.

Introduce the pair of Hermitian matrices
\begin{equation} W=G+i\Omega
\end{equation}
\begin{equation}
\lambda =(\lambda_{\alpha \beta }),\qquad
\lambda_{\alpha \beta }=\partial_i\theta_\alpha
\Lambda^{ij}\partial_j\theta_\beta \end{equation}
The next proposition provides robust criteria for
the constraint surface to be a K\"ahler manifold.

\vspace{4mm} {\bf Proposition.}
{\it With above notations and definitions the
following statements are equivalent:}

{\it i) }$(\Sigma ,\Lambda |_\Sigma )${\it \ is a K\"ahler manifold;}

{\it ii) }$({\cal M},W)${\it \ is a K\"ahler manifold;}

{\it iii) there exist a (complex) basis of constraints }$\theta _\alpha $
{\it \ and the holomorphic coordinates on }$M${\it \ in which the matrix }
$ \Pi=(\partial _i\theta _\alpha )${\it \ takes the block-diagonal form}

\begin{equation} \Pi =\left( \begin{array}{cc} \pi^{+} & 0 \\ 0 & \pi^{-}
\end{array} \right) , \end{equation}
{\it where }$\pi^{\pm }${\it \ are } $n\times m${\it \ matrices;}

{\it iv) }${\rm rank}(\lambda )=m${\it \ and}

\begin{equation} T_{ijk}=\nabla_i\omega_{jk}-g^{\alpha \beta }\nabla
_i\partial _k\theta _\beta P_j^m\partial_m\theta_\alpha -(j\leftrightarrow
k)=0, \end{equation} {\it where }$\nabla $
{\it \ is the torsion-free connection respecting metric }$g$
{\it \ and }$P_j^i=\delta _j^i+J_j^i${\it .}

\vspace{4mm}{\it Sketch of the proof}.  Since the imaginary parts of both the
Hermitian matrices $\Lambda |_\Sigma $ and $W$ are non-degenerate the first two
statements hold iff
\begin{equation} {\rm rank}\Lambda |_\Sigma =\frac 12\dim
\Sigma =n-m,\qquad {\rm rank} W=\frac 12\dim {\cal M}=n+m, \label{rankcon}
\end{equation}
provided the right and the left kernel distributions of the forms
are integrable.  The proof of equivalence of these conditions to each other and
to the algebraic parts of iii) and iv) is a simple exercise in linear algebra.

Now assuming the rank conditions (\ref{rankcon}) to be satisfied let us examine
the integrability of the almost-K\"ahler structures $W$ and $\Lambda |_\Sigma $.
As was mentioned in the previous section this is equivalent to vanishing the
respective Lichnerowicz torsion (\ref{LT}).  Consider first the case of the
extended phase-space $({\cal M},W).$ Let $D$ be the torsion-free connection
respecting metric $G=G_{AB}d\xi ^Ad\xi ^b$ and let
${\widetilde{ {\bf \Gamma}}}_{BC}^A$ be the corresponding Christoffel symbols.  Splitting the coordinates
${\xi ^A}$ as $(x^i,\eta^\alpha )$ we find only two nonvanishing components of
the Christoffel symbols,
\begin{equation}
{\widetilde{{\bf \Gamma}}}_{jk}^i=\Gamma_{jk}^i+
g^{in}\partial_n\theta _\alpha
{\widetilde{{\bf \Gamma }}}_{jk}^\alpha ,\qquad
{\widetilde{{\bf  \Gamma }}}_{ij}^\alpha
=-g^{\alpha \beta }\nabla _i\partial_j\theta _\beta ,
\label{Chr}
\end{equation} where
$\nabla =\partial +\Gamma $ is the torsion-free connection
respecting $ g$.  Then the only nonzero component of the Lichnerowicz torsion
$ T_{ABC}=\Omega _{AN}T_{BC}^N=D_A\Omega _{BC}$ is given by
\begin{eqnarray}
T_{ijk} &=&D_i\Omega _{jk}=\nabla _i\omega _{jk}+{\bf \Gamma }_{ij}^\alpha
\Omega _{\alpha k}+{\bf \Gamma }_{ik}^\alpha \Omega _{j\alpha }= \notag \\ &&
\label{tor} \\ &=&\nabla _i\omega _{jk}+{\bf \Gamma }_{ik}^\alpha P_j^m\partial
_m\theta _\alpha -(j\leftrightarrow k) \notag \end{eqnarray}
Note that for the
K\"ahler manifold $(M,\Lambda )$ the first term in rel.  ( \ref{tor}) vanishes.
Due to the rank condition (\ref{rankcon}) the form $W$ has exactly $m+n$ left
null-vectors \begin{equation} X=\xi ^i(x)\frac \partial {\partial x^i}+v^\alpha
(x)\frac \partial {\partial \eta^\alpha }, \end{equation} and $2n$ of them are
obviously of the form $\partial /\partial \eta^\alpha $ .  So, there are $n-m$
additional null-vectors of the form $Y=\xi ^i\partial _i$ with $\xi $'s
satisfying \begin{equation} \xi ^i\Lambda _{ij}=0,\qquad \xi ^i\partial _i\theta
_\alpha =0 \label{tan} \end{equation} The last equation implies that the vector
$Y$ {\it (i)} is tangent to the constraint surface $\Sigma $ and {\it (ii)} it is
annihilated by $\Lambda $ and therefore by $\Lambda |_\Sigma $.  These $n-m$
left null-vectors span a half of complexified tangent bundle $T^{{\Bbb C}}\Sigma
.$ The complimentary distribution of the right null-vectors is obtained by the
complex conjugation.  The integrability of both the distributions immediately
follows from equation (\ref{tan}) by the Frobenius theorem.$\Box $

The two nonzero components (\ref{Chr}) of the Christoffel symbols ${
\widetilde{{\bf \Gamma}} }_{BC}^A$ admit a straightforward interpretation in
terms of inner geometry of $M$.  Namely, ${\widetilde{{\bf \Gamma}} }_{jk}^i$
are the Christoffels of the unique connection ${\widetilde{{\bf \nabla}} }$
which preserves both the degenerate symmetric tensor $ \widetilde{g}^{ij}$
(the
upper left block of the inverse metric $G^{-1}$ (\ref{BM})) and its
null-covectors $d\theta _\alpha $.  Adding to ${\widetilde{{\bf \Gamma}} }$ the
torsion $T_{jk}^i=\widetilde{\omega }^{in}T_{njk}$
(\ref{tor}) we get the
covariant derivative
$\widetilde{\nabla }^{\prime}=\widetilde{{\bf \nabla }} +T$,
which is continuing to respect $\widetilde{g}^{ij}$ and $d\theta _\alpha $
and it also preserves the Dirac bi-vector $\widetilde{\omega }^{ij}$.
As to the
$M$-tensors
${\widetilde{{\bf \Gamma}} }_{ij}^{\alpha}$, $\alpha =1,...,2m,$
they can be identified with the exterior curvatures of the surface $\Sigma $
as they characterize the embedding of $\Sigma $ into the Riemann manifold
$(M,g)$.
Indeed, when $M={\Bbb R}^{2n}$ and $g$ is flat, the equations
${\widetilde{{\bf \Gamma}} }_{ij}^{\alpha}=0$ describe
$2(n-m) $-dimensional planes in
${\Bbb R}^{2n}$.

\section{Concluding remarks}

In this note we study the Hamiltonian reduction of an (almost-)K\"ahler manifold
by second-class constraints.  The criteria are found providing the reduced
dynamics to be still of a K\"ahler structure.

The main technique point of our approach is the conversion of the second-class
constraint system to a first-class one in the extended phase space ${\cal M}
=M\times {\Bbb R}^{2m}$, being a direct product of the original phase manifold
to the linear space which dimension equals to the number of the second-class
constraints.  This space is naturally endowed with the special symplectic and
metric structures invariant under translations along ${\Bbb R}^{2m}$.  It turns
out that the K\"ahler geometry of the reduced phase space (being given
implicitly until one keeps the constraints unresolved) is completely encoded,
and may be studied, by means of the explicit K\"ahler structure on the extended
phase space ${\cal M}$.

Although ${\cal M}$ has appeared as a suitable auxiliary construction to study
the classical Hamiltonian reduction its role becomes even more important upon
the quantization.  Let us take a quick look at this point.  When ${\cal M}$ is
the K\"ahler manifold it can be easily quantised (applying the construction
of the paper \cite{DLS}) to produce the algebra of Wick
symbols being an associative deformation in $ \hbar $ of the ordinary
commutative algebra of functions on ${\cal M}$.  In so doing, the invariance of
the K\"ahler structure under the $ {\Bbb R}^{2m}$-shifts is transformed into
the respective symmetry of the Wick star-product.  As the result,
the functions,
being constant on ${\Bbb R}^{2m}$, form a closed subalgebra, and thus the Wick
star-product on ${\cal M}$ induces that on the original phase-space $M$.  The
latter star product is characterized by the nontrivial central subalgebra
generated by the second-class constraints (which are Casimir functions of the
respective Dirac bracket) and it is compatible with the Wick polarization.  It
is almost obvious that the classical limit of the corresponding star-commutator
on $M$ would be nothing but the classical Dirac bracket.  In such a manner, the
deformation quantization on ${\cal M}$ gives rise to what one may naturally call
as {\it quantum Dirac brackets of Wick type}.

\vspace{0.5cm}

\noindent {\bf Acknowledgements.}
The authors appreciate the partial financial
support from RFBR under the grant no 00-02-17-956, INTAS under the grant no
00-262 and Russian Ministry of Education under the grant E-00-33-184.  The work
of AAS is partially supported by RFBR grant for support of young scientists no
01-02-06420.  SLL apreciates support from STINT and the warm hospitality of the
Chalmers University of Technology where this manuscript has been finished.

\end{document}